\documentclass[twocolumn]{jpsj3}
\usepackage{dcolumn}
\usepackage{bm}
\usepackage{amsmath}
\usepackage{txfonts}
\usepackage[T1]{fontenc}
\usepackage{xspace}
\usepackage{ulem}
\usepackage{comment}
\usepackage{braket}
\ifx\pdfoutput\undefined
\usepackage[dvipdfmx]{graphicx}
\usepackage[dvipdfmx]{hyperref}
\usepackage[dvipdfmx]{xcolor}
\else
\usepackage{graphicx}
\usepackage{hyperref}
\usepackage{color}
\fi

\makeatletter
\providecommand{\ext@figure}{lof}
\providecommand{\ext@table}{lot}
\makeatother

\let\emph\textit

\title{
  Stealthy hyperuniform disorder: A new route to
  controlling electronic states and magnetic properties
  in correlated systems
  }

\author{Akihisa \textsc{Koga}$^{1,2}$ and Takanori \textsc{Sugimoto}$^{3,4,5,6}$}

\inst{
  $^1$Department of Physics, Chuo University, Bunkyo, Tokyo 112-8551, Japan\\
  $^2$Department of Physics, Institute of Science Tokyo, Meguro, Tokyo 152-8551, Japan\\
  $^3$Department of Electrical, Electronic and Information Engineering, Kansai University, Suita, Osaka 564-8680, Japan\\
  $^4$Center for Quantum Information and Quantum Biology, Osaka University, Toyonaka, Osaka 560-0043, Japan\\
  $^5$Advanced Science Research Center, Japan Atomic Energy Agency, Tokai, Ibaraki 319-1195, Japan\\
  $^6$Computational Materials Science Research Team, Riken Center for Computational Science (R-CCS), Kobe, Hyogo 650-0047, Japan
}

\date{\today}
\abst{
  We investigate how spatial correlations in bond disorder affect the
  electronic and magnetic properties of the Hubbard model on the honeycomb lattice.
  Hyperuniform structures, distinct from random and quasiperiodic ones,
  have recently attracted considerable interest due to
  their anomalous suppression of density fluctuations.
  By diagonalizing the noninteracting Hamiltonian,
  we show that a linear density of states (DOS) robustly emerges,
  while the stealth property of the bond distribution
  changes the wave functions in the higher-energy region extended and
  significantly modifies the DOS near the band edge.
  To clarify the impact on magnetism,
  we apply the real-space Hartree approximation to the Hubbard model.
  We find that,
  the phase transition always occurs between semimetallic and antiferromagnetically
  ordered states and
  the staggered magnetization in the ordered state is sensitive to
  the stealth property.
  A comparison with the quasiperiodic honeycomb tiling further highlights
  the role of structural correlations.
  These results demonstrate that stealthy hyperuniform disorder
  provides a novel route to controlling electronic states and
  magnetic properties in correlated systems.
}

\begin{document}
\maketitle

\section{Introduction}
Recently, point distributions have attracted much attention
since disordered hyperuniform structures~\cite{Torquato_2003,Torquato_2016}
have been observed in various fields
such as soft matter~\cite{Berthier_2011,Kurita_2011},
solids~\cite{Zheng_2020,Llorens_2020}, active matter~\cite{Huang_2021},
biology~\cite{Jiao_2014}, and astronomy~\cite{Philcox_2023}.
A hyperuniform structure is defined as a distribution
in which density fluctuations are anomalously suppressed
at large length scales, i.e., the structure factor 
$S({\bf k})$ tends to zero as ${\bf k}\rightarrow 0$,
whereas typical random (non-hyperuniform) distributions exhibit
finite values in this limit.
Among hyperuniform systems, stealthy hyperuniform structures --
characterized by a vanishing structure factor within a finite range
around the origin in reciprocal space --
have attracted particular interest; for example,
in photonic crystals~\cite{Man_2013,Asakura},
the stealthness of the point configurations has been suggested
to enhance the photonic band gap.
This example motivates us to ask whether hyperuniform structural correlations
can also affect electronic wave functions and magnetic ordering in quantum
lattice systems.

While stealthy hyperuniformity has been extensively studied
at the level of point distributions~\cite{Koga_Sakai_2024},
its implications for lattice models relevant to
condensed-matter systems remain less explored.
To investigate hyperuniformity in a lattice setting,
it is important to employ a unified framework that
treats aperiodic systems on equal footing.
One of the models that can simultaneously incorporate periodicity,
quasiperiodicity, and disorder is the honeycomb lattice with
two types of bonds~\cite{KogaMatsubara}.
By appropriately arranging the bond distributions,
this model allows for a systematic comparison of different types of structural order.
Previous studies based on the Hubbard model on the honeycomb lattice mainly examined
how randomness and quasiperiodicity in the bond distribution influence
the density of states and magnetic phase transitions~\cite{KogaMatsubara}.
However, the role of stealthiness of bond distributions has not yet been clarified.
The central question addressed in this work is how spatial correlations in bond
disorder, beyond its amplitude alone, affect electronic states and magnetic
ordering in the Hubbard model.
Stealthy hyperuniform bond distributions provide a useful platform for addressing
this question since they suppress long-wavelength fluctuations of the bond
distribution without relying on conventional periodicity.
This comparison allows us to distinguish the effects of stealthy correlations
from those of random and quasiperiodic bond distributions.

The paper is organized as follows.
In Sec.~\ref{sec:model},
we introduce the model Hamiltonian and construct bond distributions
with stealthy hyperuniform structures.
In Sec.~\ref{sec:results},
we examine their effects on the noninteracting density of states and
clarify how the stealth property modifies the wave functions near the band edge.
We then analyze magnetic properties within
the real-space Hartree approximation and
compare the results for the systems with random, stealthy hyperuniform, and
quasiperiodic bond distributions. 
Finally, a summary is presented in Sec.~\ref{sec:summary}.

\section{Model and method}\label{sec:model}

\subsection{Correlated electron systems on honeycomb lattice}
\begin{figure}[htb]
  \includegraphics[width=\linewidth]{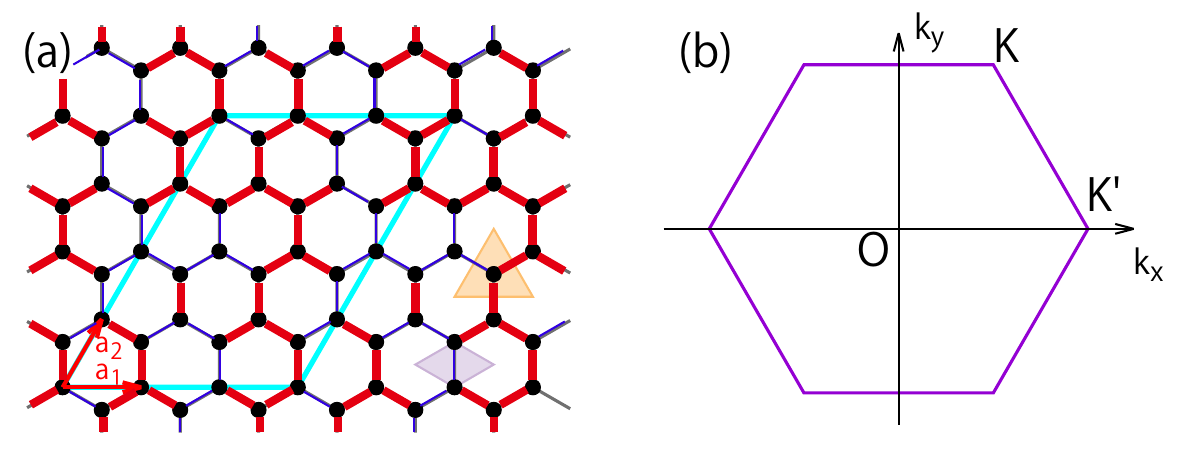}
  \caption{
    (a) Honeycomb lattice and its unit vectors ${\bf a}_1$ and ${\bf a}_2$.
    Bold red and thin blue lines represent $\alpha$- and $\beta$-bonds, respectively.
    Parallelogram represents the system specified
    by $(L_1, L_2)=(3,4)$.
    Shaded triangle and rhombus indicate the areas associated with
    one site and one bond, respectively.
    (b) First Brillouin zone of the honeycomb lattice.
  }
  \label{model}
\end{figure}

First, we explain geometric properties
of the honeycomb lattice.
Its unit vectors are given by
${\bf a}_1=(\sqrt{3},0)a$, and ${\bf a}_2=(\sqrt{3}/2,3/2)a$,
where $a$ denotes the distance between nearest-neighbor sites,
as shown in Fig.~\ref{model}(a).
We treat a finite-size system consisting of $N=2L_1L_2$ sites, with integers 
$L_1$ and $L_2$, under periodic boundary conditions.
In this case, the wave vector is given by
${\bf k}=(m_1/L_1){\bf b}_1+(m_2/L_2){\bf b}_2$,
where ${\bf b}_1=2\pi(1/\sqrt{3}, -1/3)/a$, ${\bf b}_2=2\pi(0,2/3)/a$,
and $m_1$ and $m_2$ are integers.
The first Brillouin zone of the honeycomb lattice
is shown in Fig.~\ref{model}(b).
Its corners located at $(2\pi/(3\sqrt{3}a),2\pi/(3a))$ and
$(4\pi/(3\sqrt{3}a),0)$ are known as
the K and K' points, respectively.

In this study, we consider correlated electron systems
to discuss how the spatial distribution of two kinds of bonds affects
their ground state properties.
To this end, we treat the Hubbard model on the honeycomb lattice,
whose Hamiltonian is given as
\begin{align}
  H&=-t_\alpha\sum_{\langle ij\rangle_\alpha}^{N_{B\alpha}}c_{i\sigma}^\dag c_{j\sigma}
  -t_\beta\sum_{\langle ij\rangle_\beta}^{N_{B\beta}}c_{i\sigma}^\dag c_{j\sigma}
  +U\sum_i n_{i\uparrow} n_{i\downarrow},
\end{align}
where $\langle ij\rangle_\gamma$ stands for the nearest-neighbor pair
on the $\gamma(=\alpha, \beta)$ bonds, and
$N_{B\gamma}$ is the number of $\gamma$-bonds.
$c_{i\sigma} (c^\dag_{i\sigma})$ annihilates (creates) an electron
at the $i$th site with spin $\sigma$.
$t_\gamma$ is the hopping integral for the $\gamma$-bond
and $U$ is the onsite Coulomb interaction.
Since the honeycomb lattice is bipartite,
the chemical potential is always $\mu=U/2$
when the electron density is fixed to be half filling.
When $t_\alpha=t_\beta$, the system reduces to the Hubbard model
on the regular honeycomb lattice.
It is known that, in the noninteracting case ($U=0$),
the Dirac-type dispersion appears around the K and K' points.
Owing to the lack of the density of states (DOS) at the Fermi level,
the introduction of the Coulomb interaction does not immediately
induce magnetic order.
Consequently, the quantum phase transition to the antiferromagnetically ordered state
occurs at a finite interaction strength,
which has been discussed so far~\cite{Sorella_1992,Furukawa_2001,Feldner_2010,Sorella_2012,Assaad_2013,Raczkowski_2020,Ostmeyer_2020,Ostmeyer_2021}.

\subsection{Stealthy hyperuniform bond distribution}

The Hubbard model with quasiperiodic and randomly distributed bond structures
has been examined in Ref.~\cite{KogaMatsubara},
where significant changes in the DOS and
the critical interaction strengths were reported.
However, the role of the disordered structure itself -- particularly
the case of stealthy hyperuniform bond distributions -- has not been examined.
It is instructive to discuss how distinct bond distributions affect
the ground state within a unified framework.

For this purpose, we characterize the bond distribution
through its structure factor.
Here, we define the coordinate of each bond 
as the center of the two sites it connects.
Its structure factor $S_\gamma({\bf k})$ is defined as,
\begin{align}
  S_\gamma({\bf k})&=\frac{1}{N_B}\big|\, \rho_\gamma({\bf k}) \,\big|^2,\\
  \rho_\gamma({\bf k})&=\sum_ie^{i{\bf k}\cdot{\bf r}^{(\gamma)}_i},
\end{align}
where ${\bf r}_i^{(\gamma)}$ is the coordinate of the $i$th $\gamma$-bond
and $N_B(=\sum_\gamma N_{B\gamma})$ is the total number of bonds.
In the following, we neglect the trivial sum at the $\Gamma$ point.
Since the $\alpha$- and $\beta$-bond configurations are complementary,
$\rho_{\alpha}({\bf k})+\rho_{\beta}({\bf k})=0$ for
${\bf k}\neq 0$. This gives
$\rho_{\alpha}({\bf k})=-\rho_{\beta}({\bf k})$, and consequently
$S_{\alpha}({\bf k})=S_{\beta}({\bf k})$. 
We thus omit the bond index of the structure factor.
When $S({\bf k})$ tends to zero as ${\bf k}\rightarrow 0$,
we refer to the bond distribution as hyperuniform.
Such a bond distribution is characterized by
the suppression of long-wavelength fluctuations.
This is in contrast to the uncorrelated random bond distribution,
where $S({\bf k})$ takes a constant value independent of ${\bf k}$.

\begin{figure*}[htb]
  \includegraphics[width=\linewidth]{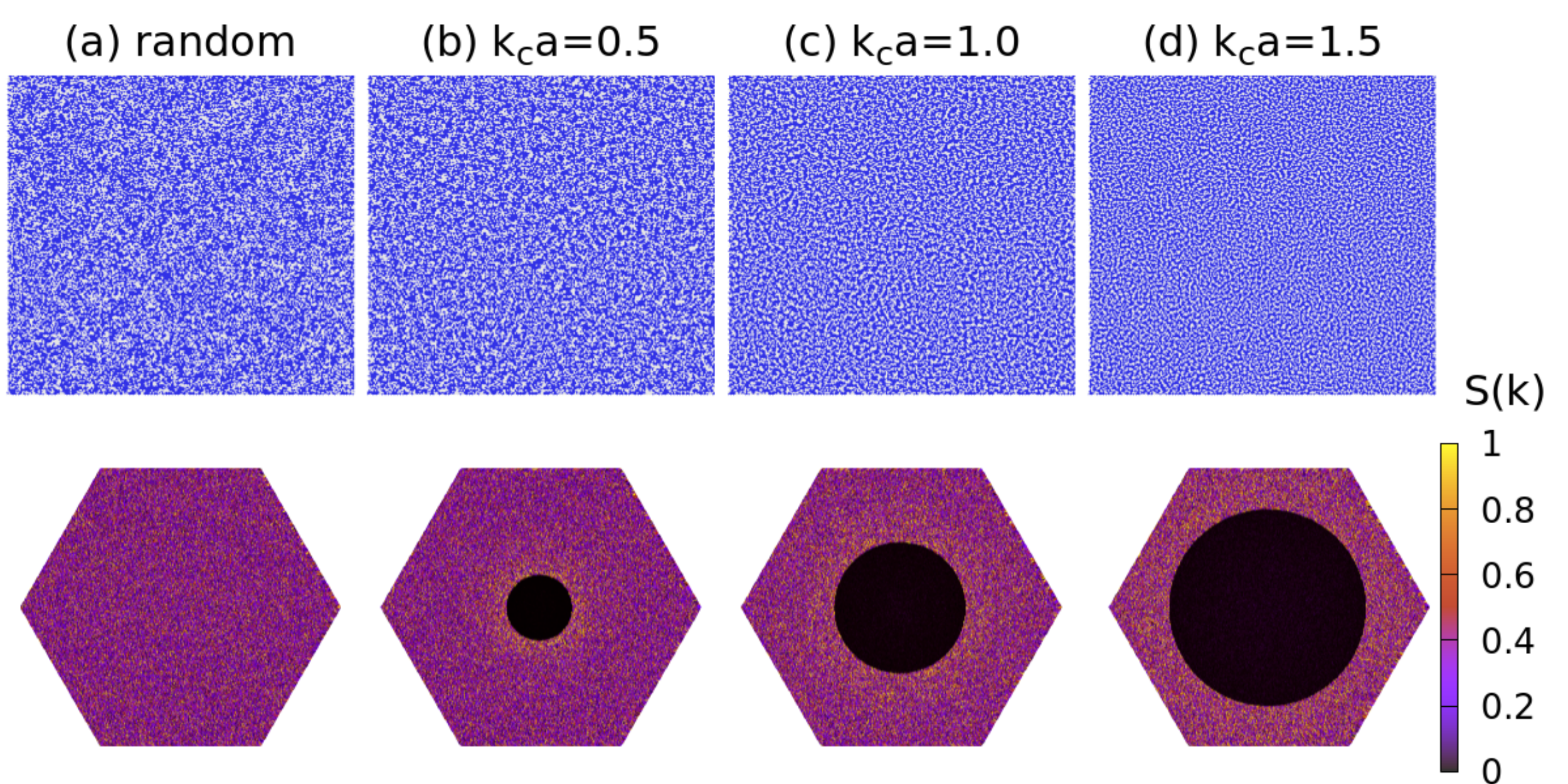}
  \caption{
    Upper panels show the distributions of the bonds
    for the honeycomb systems with (a) $k_ca=0.0$, (b) 0.5, (c) 1.0,
    and (d) $1.5$.
    $\alpha$- and $\beta$-bonds are represented by light gray and blue rhombi,
    respectively 
    (see also Fig.~\ref{model}).
    Lower panels show the corresponding structure factors.
  }
  \label{fig:lattice}
\end{figure*}
In the study, we set the number ratio between two bonds as
$N_{B\alpha}/N_{B\beta}=\tau$, for later convenience,
where $\tau[=(1+\sqrt{5})/2]$ is the golden ratio.
To obtain the bond distribution
which satisfies the stealthy hyperuniform structure,
we use the reverse Monte Carlo method.
Here, we introduce the cost function
\begin{align}
  \Phi\left(\{{\bf r}^{(\alpha)}\}\right)=\sum_{\bf k} V({\bf k})\left[\, S({\bf k})-S_{\rm target}({\bf k})\,\right]^2,
  \label{cost}
\end{align}
where $V({\bf k})$ is the window function,
and $S_{\rm target}({\bf k})$ is the target structure factor.
In this work, we set $V({\bf k})=1$ for $0<|{\bf k}|<k_c$ and
$V({\bf k})=0$ otherwise,
where $k_c$ is some positive number.
In the optimization procedure, two bonds are randomly selected and swapped, and
the update is accepted only if the cost function is reduced.
By repeating this procedure, the bond distribution is gradually optimized so that
$S({\bf k})$ approaches $S_{\rm target}({\bf k})$.
We iterate this scheme until the cost function satisfies $\Phi/N_c<0.01$,
where $N_c$ is the number of ${\bf k}$ points satisfying $V({\bf k})=1$.
For the stealthy hyperuniform distribution,
we set $S_{\rm target}({\bf k}) = 0$, so that $S({\bf k})$ is suppressed for $0 < |{\bf k}| < k_c$.
Thus, fluctuations of the bond distribution are suppressed on length scales longer than about $2\pi/k_c$.
In other words, $k_c$ controls the length scale above which the bond distribution becomes homogeneous:
$k_c = 0$ corresponds to the random distribution, and with increasing $k_c$,
the fluctuations are suppressed down to shorter length scales.
The number $N_c$ increases with $k_c$,
while the number of degrees of freedom in the bond configuration is fixed.
Therefore, the stealth condition is expected to become difficult to satisfy for large $k_c$.
In this study, we focus on the range $k_c a \leq 1.5$,
where the condition $\Phi/N_c<0.01$ is satisfied.
We note that the cost function Eq.~(\ref{cost}) is used only to generate the bond
distribution and is not a part of the Hubbard Hamiltonian.
The obtained bond distribution is then incorporated into the Hubbard
Hamiltonian by assigning the corresponding hopping amplitudes to the selected
nearest-neighbor bonds.

Applying the reverse Monte Carlo method to the system with randomly
distributed bonds, we obtain stealthy hyperuniform bond distributions.
Figure~\ref{fig:lattice} shows parts of the honeycomb lattices and their structure
factors for the system with $L_1 = L_2 = 512$.
In the random case
[Fig.~\ref{fig:lattice}(a)], clusters of $\alpha$-bonds with various sizes are distributed
inhomogeneously.
As $k_c$ increases, the clusters become smaller and more
uniform in size, and the large-scale inhomogeneity disappears
[Figs. \ref{fig:lattice}(b)-\ref{fig:lattice}(d)].
This reflects the suppression of the fluctuations for $k < k_c$,
which is clearly seen in the corresponding structure factors.

Although the stealth constraint primarily affects the long-range
behavior of the bond distribution, local properties are also influenced
by the choice of $k_c$.
\begin{figure}[htb]
  \includegraphics[width=\linewidth]{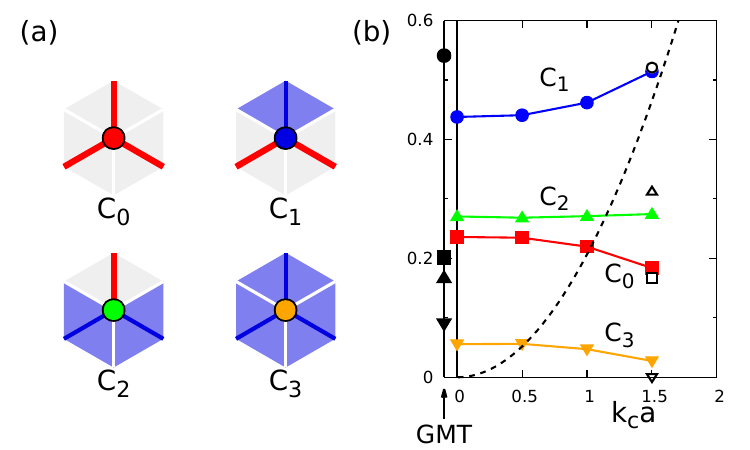}
  \caption{
    (a) Four kinds of vertices in the honeycomb lattice with disordered bonds.
    C$_i$ vertex is connected to the nearest-neighbor vertices
    by $(3-i)$ $\alpha$-bonds (bold red lines) and
    $i$ $\beta$-bonds (thin blue lines),
    which are specified by the hoppings $t_\alpha$ and $t_\beta$.
    (b) Fractions of four vertices as a function of $k_c$.
    The dashed line represents the ratio $\pi k_c^2/S_{\rm BZ}$ with the square of the first Brillouin zone $S_{\rm BZ}$.
    Solid and open symbols represent the vertex fractions for the golden-mean honeycomb tilings~\cite{KogaMatsubara}
    and additionally constrained stealthy hyperuniform point pattern with $k_ca=1.5$ (see text).
  }
  \label{fig:vertex}
\end{figure}
Two kinds of bonds are introduced in the honeycomb lattice,
and four kinds of vertices C$_0$, C$_1$, C$_2$, and C$_3$ appear,
as shown in Fig.~\ref{fig:vertex}(a).
A C$_i$ vertex is connected to the nearest-neighbor vertices by $(3-i)$ $\alpha$-bonds
and $i$ $\beta$-bonds.
Their fractions are shown in Fig.~\ref{fig:vertex}(b).
The results are obtained from nine independent examples with $N=524,288$ and
their statistical error is smaller than the size of symbols.
When $k_ca=0.5$, the fractions are almost the same as
those for the random case ($k_c=0$).
This means that the imposed constraint is insufficient to produce a bond distribution
which differs significantly from the random case.
By contrast, in the case with $k_ca=1.5$, the window function covers
about half of the first Brillouin zone
and such a strong constraint significantly affects
the spatial distributions of bonds and vertices.
In fact, this reduces the fractions for C$_0$ and C$_3$ vertices,
which correspond to locally homogeneous bond environments.

For comparison, we show the vertex fractions 
in the quasiperiodic deterministic honeycomb tiling~\cite{KogaMatsubara}.
This tiling is composed of five kinds of hexagonal tiles
and contains two characteristic length scales $\ell$ and $s(=\ell/\tau)$.
Since this tiling is topologically equivalent to the regular honeycomb lattice,
we can regard it as the regular honeycomb lattice
with two kinds of bonds $(\alpha, \beta)$ modulated quasiperiodically.
The number ratio of bonds is given by $N_{B\alpha}/N_{B\beta}=\tau$.
We find in Fig.~\ref{fig:vertex}(b)
the vertex fractions differ from those in the systems with
stealthy hyperuniform bonds.
Therefore, we can discuss how the ground state properties are affected by
the bond distributions as well as the vertex fractions.
In the following, we fix the hopping ratio $t_\alpha=2t_\beta$,
take $t=t_\beta$ as the unit of energy.
We expect that the qualitative behavior discussed below does not depend
on the specific value of the ratio as long as $t_\alpha \ne t_\beta$,
since the ratio mainly sets the energy scale of the difference between
the two bonds.

\section{Results}\label{sec:results}

\begin{figure}[htb]
  \includegraphics[width=\linewidth]{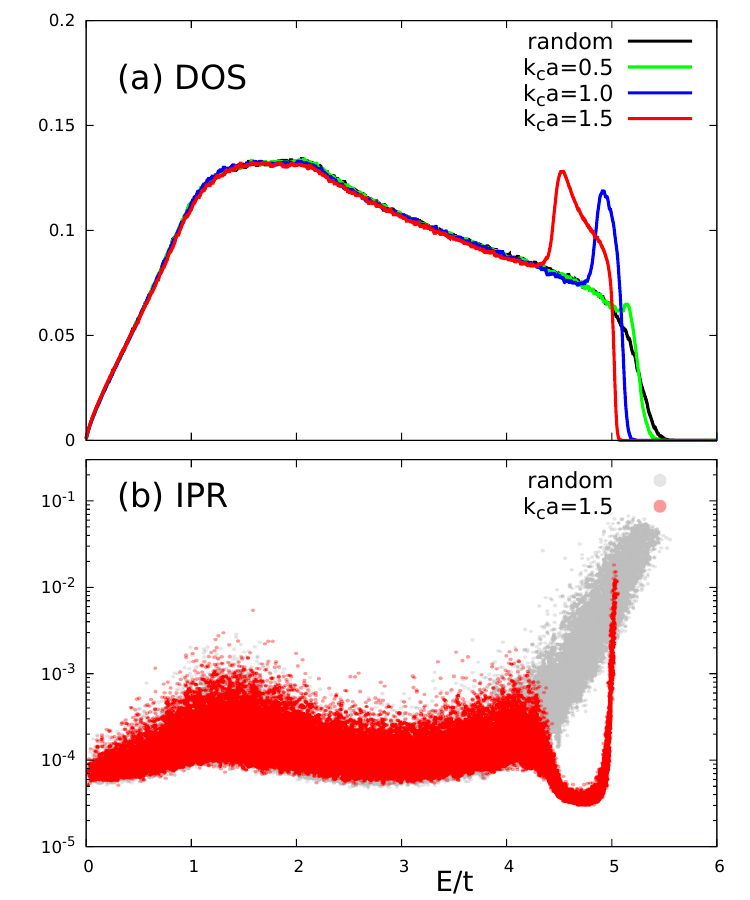}
  \caption{
    (a) Density of states for the tight-binding model
    with $k_ca=0, 0.5, 1.0$, and $1.5$
    when $L_1=L_2=512$.
    (b) Logarithmic plot of the IPR spectra
    for the model with $k_ca=0$ and $1.5$ when $L_1=L_2=256$.
  }
  \label{dos}
\end{figure}

\subsection{Effect of the stealthy hyperuniform patterns
  on noninteracting system}
Here, we consider the tight-binding model with disordered bonds.
Diagonalizing the Hamiltonian with $U=0$,
we obtain the density of states (DOS), which is defined as
\begin{align}
  D(E)&=\frac{1}{N}\sum_i\delta(E-E_i),
\end{align}
where $E_i$ is the $i$th eigenvalue of the Hamiltonian
and $N(=2L_1L_2)$ is the number of sites.
The DOS for the system with $L_1=L_2=512$
is shown in Fig.~\ref{dos}(a).
In the case, we have replaced the delta function with
the Gaussian $\delta(E)\sim \exp( -(E/\epsilon)^2 )/(\sqrt{\pi}\epsilon)$
with the width $\epsilon=0.01t$.
We find that
the linear DOS is always observed around $E=0$
and its slope is little affected by $k_c$.
Moreover, the curve of the DOS for $|E/t|<4$ is almost identical,
indicating that the stealthy properties have little effect on
this energy region and
the DOS mainly depends on the ratio $N_{B\alpha}/N_{B\beta}$.
By contrast,
the DOS around the band edge $|E/t|\sim 5$
shows a strong dependence on $k_c$.
For $k_c=0$, the system has randomly distributed bonds
and the DOS smoothly decreases up to $|E/t|\sim 5.5$.
As $k_c$ increases, the band edge shifts downward and
the DOS just below the band edge increases.
In particular, in the case with $k_ca=1.5$,
the band edge is located around $E/t\sim 5$,
which is approximately ten percent smaller than
that for the random bond case with $k_c=0$.
The change of the DOS at high energies can be understood as follows.
For an eigenstate with energy $E$, the Schr\"odinger equation at the 
site $j$ with the largest amplitude gives $|E| \le \sum_k t_{jk}$, where the
sum runs over the three bonds connected to that site. For
$t_\alpha/t_\beta = 2$, this sum exceeds $5t$ only for the
C$_0$ vertices, and thus the states with $|E| > 5t$ are
supported by the regions where the $\alpha$-bonds are concentrated.
Since such regions are suppressed with increasing $k_c$ (Fig.~\ref{fig:vertex}),
the band edge shifts downward [Fig.~\ref{dos}(a)].

To clarify how the stealthness in the bond distributions affects
localization properties of the one-particle wave functions,
we examine the inverse participation ratio (IPR) of the 
eigenfunction, which is defined as
\begin{align}
  {\rm IPR}=\sum_{{\bf r}_i} \big|\psi({\bf r}_i)\big|^4,
\end{align}
where $\psi$ is the normalized eigenfunction of the tight-binding Hamiltonian.
For a wave function localized on a few sites,
the IPR takes a relatively large value,
whereas for an extended state it becomes small and
scales inversely with the system size.
In Fig.~\ref{dos}(b), we show 
the IPR for the disordered systems with $L_1=L_2=256$.
In the low energy region $|E|/t<4$,
the IPR values are distributed within ${\rm IPR}\lesssim 0.003$
and are almost identical
between two cases $k_ca=0$ and $1.5$.
This is consistent with the fact that
the DOS is not affected by the stealthness of the bond distributions
discussed above.
On the other hand, a fairly large difference appears
in the high energy region.
In the random case ($k_c=0$),
the IPR is widely scattered and gradually increases
as the energy approaches the band edge,
indicating that the wave function in high energy region
tends to become localized.
In the stealthy hyperuniform case with $k_ca=1.5$,
similar properties appear in the vicinity of the band edge
although
the IPR values collapse almost onto a single curve and
the rapid increase appears over a narrower energy window.
The most important point is that,
in the energy region $(4.2\lesssim |E|/t \lesssim4.8)$ 
where peak structures appear in DOS,
the IPR takes its smallest values.
This demonstrates that the introduction of stealthness strongly suppresses
localization and makes the wave functions significantly more extended.
This can be understood from the argument above: in the random case, the
high-energy states tend to be localized in the regions where the
$\alpha$-bonds are concentrated, and the suppression of such regions
with increasing $k_c$ makes these states more extended.
These findings highlight that stealthness in the structure factor
has a pronounced impact on the nature of the wave functions,
leading to substantial modifications in the high-energy spectrum of the system.
Similar delocalization behavior has also been reported
in one-dimensional hyperuniform systems~\cite{Jeon_2026,Vanoni_2026}.

The stealth property is imposed near the $\Gamma$ point in reciprocal space,
well separated from the K and K' points.
Therefore, one may naively expect that this property
primarily modifies the high-energy sector of the noninteracting
Hamiltonian, which is consistent with the above results.
By contrast, if the stealth constraint were imposed in the
vicinity of the K and K' points, one would expect it to influence
the low-energy properties since these points govern the
Dirac-like excitations.

\begin{figure}[htb]
  \includegraphics[width=\linewidth]{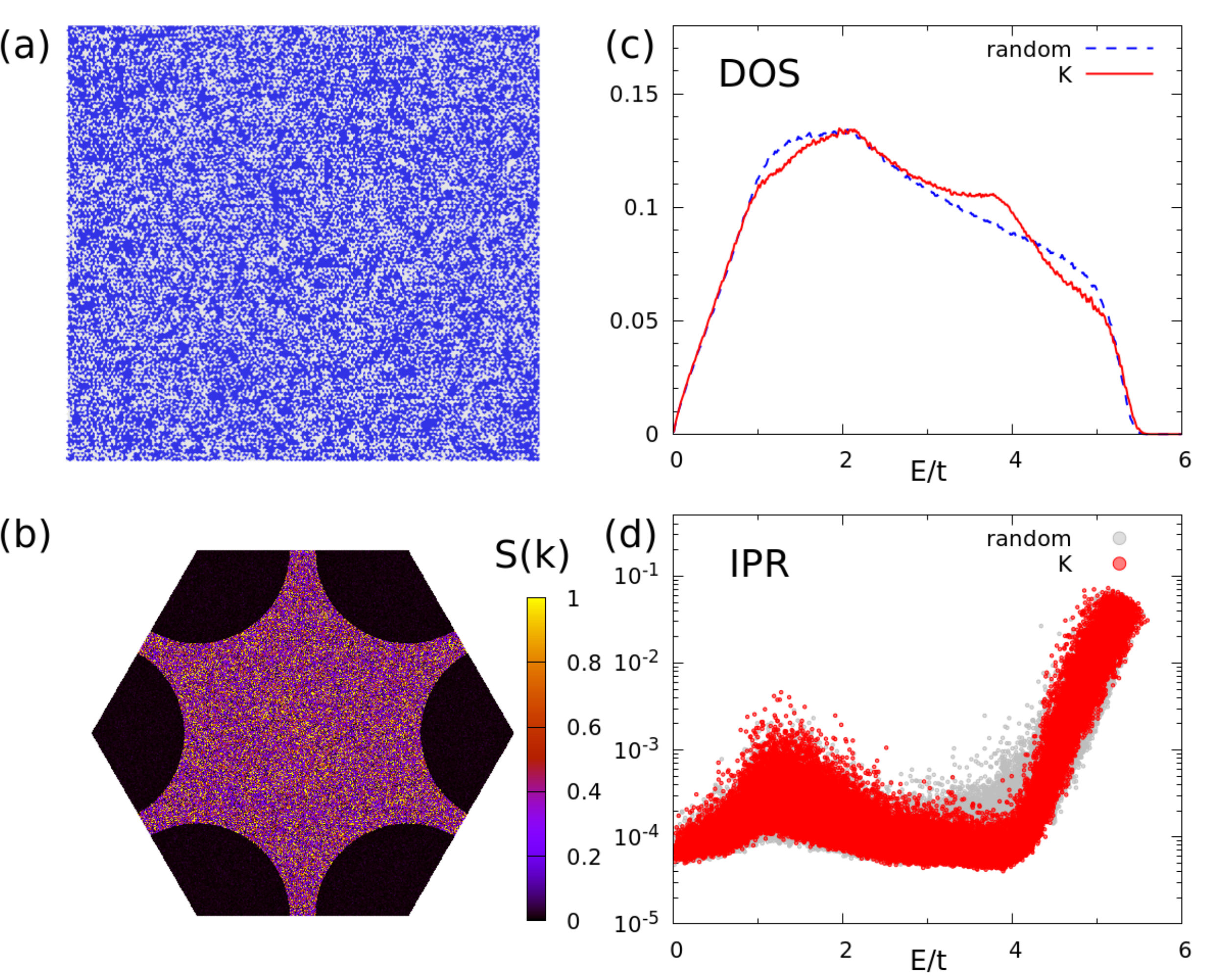}
  \caption{
    (a) Bond distribution for the K-pattern with $k_ca=1.5/\sqrt{2}$
    and (b) its structure factor.
    (c) DOS for the tight-binding models on the K-pattern with $L_1=L_2=512$.
    (d) IPR spectrum for the model with $L_1=L_2=256$.
    }
  \label{K}
\end{figure}

To examine this possibility,
we consider the window function $V({\bf k})$ defined as
$V({\bf k})=1$ within a circular region with radius $k_c$
around the K and K' points,
and $V({\bf k})=0$ elsewhere.
This construction allows us to isolate the effect of stealthiness
imposed near the Dirac points from that originating at the $\Gamma$ point.
The bond distribution obtained by imposing this constraint
is referred to as the K-pattern.
Figure~\ref{K} shows the bond configurations for the K-pattern
with $k_ca=1.5/\sqrt{2}$,
its structure factor, the DOS for the tight-binding model,
and its IPR.
Here, $k_c a = 1.5/\sqrt{2}$ is chosen so that the total area of the
constrained region in reciprocal space is the same as that for the
$\Gamma$-pattern with $k_c a = 1.5$.
Compared with the randomly distributed system (see Fig.~\ref{fig:lattice}),
the structure factor is distinct, even though
this difference is hardly visible in real-space bond distributions.
This is because the
constraint is imposed only around the K and K' points, and the
long-wavelength fluctuations of the bond distribution remain.
We find that the low- and high-energy parts of
the DOS closely resemble that of a random system.
The similarity at high energies can be understood from the argument above:
since the long-wavelength fluctuations remain, the regions where
the $\alpha$-bonds are concentrated are not suppressed in the K-pattern.
Only in the intermediate range $1\lesssim E/t \lesssim 5$,
the slight changes are observed in both DOS and IPR.
These observations indicate that the $\Gamma$-pattern constraint produces a pronounced modification of the DOS around the band edge, while both the $\Gamma$- and K-pattern constraints affect the low-energy DOS relevant to the magnetic instability. 

\subsection{Hubbard model with disordered bonds}
Next, we examine magnetic properties of the Hubbard model.
In Hubbard-type models, interaction-driven magnetic instabilities are generally
sensitive to the low-energy electronic states near the Fermi level, although the
quantitative details depend on the lattice geometry~\cite{Hirsch1985,KogaMatsubara}.
To clarify how the bond distributions affect the competition between
the semi-metallic and antiferromagnetically ordered states,
it is crucial to treat sufficiently large systems.
Here, we employ the real-space Hartree approximation.
This method is too simple to provide an accurate
estimate of the critical interaction strength,
while it captures the essential features of the phase transition.
In fact, for the regular honeycomb lattice,
the method yields a finite critical interaction;
however, the resulting value $U_c/t\sim 2.23$~\cite{Sorella_1992} is smaller than
that obtained by the Monte Carlo method, which gives $U_c/t\sim 3.835$~\cite{Ostmeyer_2021}.
Thus, the Hartree approximation overestimates the tendency toward magnetic ordering.
Despite this quantitative discrepancy, we can still
discuss the qualitative influence of bond distributions on $U_c$.

In the real-space Hartree approximation,
we introduce the site- and spin-dependent mean-fields
$\langle n_{i\sigma}\rangle$.
The effective Hamiltonian is then obtained as
\begin{align}  
  H=-t_\alpha\sum_{\langle ij\rangle_\alpha}c_{i\sigma}^\dag c_{j\sigma}
  -t_\beta\sum_{\langle ij\rangle_\beta}c_{i\sigma}^\dag c_{j\sigma}
  +U\sum_{i\sigma} n_{i\sigma}\langle n_{i\bar{\sigma}}\rangle.
\end{align}
For given mean fields, we numerically diagonalize
the mean-field Hamiltonian, update the mean fields, and
iterate this self-consistent procedure
until the result converges within numerical accuracy.
We apply the real-space Hartree approximation to the Hubbard model
on the honeycomb lattice with $L_1=L_2=120$
under periodic boundary conditions.
In the following, we focus on the random configuration with $k_ca=0$
and the stealthy hyperuniform configuration with $k_ca=1.5$.
These distributions are slightly different from each other,
as shown in Fig.~\ref{fig:lattice},
while
we do not observe clear differences
in the spatial patterns of the local magnetizations
$m_i=(\langle n_{i\uparrow}\rangle - \langle n_{i\downarrow}\rangle)/2$.

\begin{figure}[htb]
  \includegraphics[width=\linewidth]{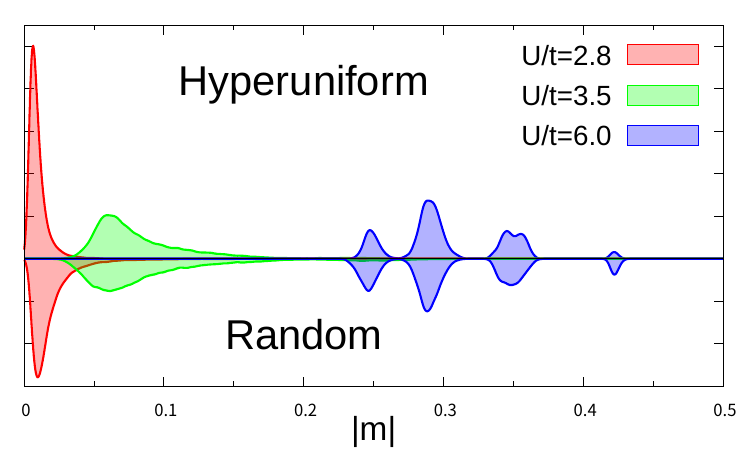}
  \caption{
    Distribution of the local magnetizations
    in Hubbard model on the honeycomb lattice with
    (upper panel) hyperuniformly and (lower panel) randomly
    distributed bonds for $U/t=2.8$, $3.5$, and $6.0$.
  }
  \label{mag2}
\end{figure}
Figure~\ref{mag2} shows the distributions of the local magnetizations
for $U/t=2.8, 3.5$ and $6.0$.
In the strong coupling case with $U/t=6.0$,
the magnetizations are classified into four groups
corresponding to different vertex types,
ordered from smaller to larger values as C$_0$, C$_1$, $\cdots$.
In particular, large magnetizations $|m|\sim 0.42$ appear
at the C$_3$ vertices
since these vertices, which are connected by three weaker $\beta$-bonds,
are more easily magnetized.
This observation is consistent with the fact that
the weight for the magnetization peak around $|m|\sim 0.42$
in the stealthy hyperuniform case
is relatively small, reflecting its smaller fraction of C$_3$ vertices
[see Fig.~\ref{fig:vertex}(b)].
As the interaction strength decreases,
the distinction among vertex types is smeared out, and 
each distribution exhibits a single peak.
We find that, at $U/t=2.8$,
the peak position for the stealthy hyperuniform case is slightly
lower than that of the random case,
indicating that the magnetically ordered state is less stable
in the system with stealthy hyperuniform bonds.

\begin{figure}[htb]
  \includegraphics[width=\linewidth]{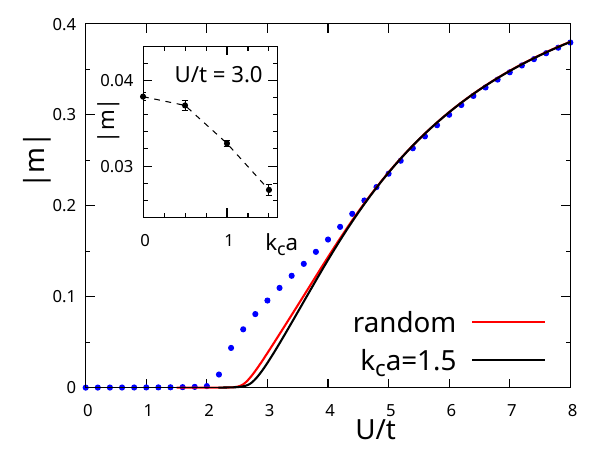}
  \caption{
    Average of local magnetizations
    as a function of the Coulomb interaction $U/t$
    in the disordered system with $N=2L_1L_2$ with $L_1=L_2=120$.
    Dotted line indicates the results for the golden-mean tilings.
    The inset shows the $k_c$ dependence of the averaged local magnetizations
    in the system with $U/t=3.0$, including the results for $k_ca=0.5, 1.0$, and $1.5$.
  }
  \label{mag}
\end{figure}
To clarify this tendency,
we calculate the spatially averaged staggered magnetization
in both disordered systems, which is shown in Fig.~\ref{mag}.
We find that
no finite magnetization appears at any site when $U < U_c$.
Thus, in the weak-coupling regime,
a nonmagnetic semi-metallic state is stabilized.
As the interaction strength increases beyond $U_c$,
the magnetization is induced and
a magnetically ordered state emerges,
as shown in Fig.~\ref{mag}.
Therefore, even in the presence of bond disorder,
a quantum phase transition occurs between the normal semimetallic and
antiferromagnetically ordered states.
We find that, when $2.5\lesssim U/t \lesssim 4.0$,
the spatially averaged magnetization for the stealthy hyperuniform system
is always smaller than that for the random case.
Furthermore, we clarify that 
the averaged staggered magnetization decreases with increasing $k_c$,
as shown in the inset of Fig.~\ref{mag}.  
These observations suggest that the critical interaction for 
the stealthy hyperuniform system may be larger than that for the random case.
In inhomogeneous systems, the magnitudes of the local magnetizations
near the critical interaction span a broad range,
from values very close to zero to relatively larger ones.
Furthermore, the magnetization profile strongly depends on the bond configurations.
These features make it difficult to determine the critical interaction precisely. 
Such behavior may also be reminiscent of rare-region effects; however, 
establishing a connection to Griffiths physics requires analyses beyond the present mean-field treatment.

We now turn to an analysis of the fine spatial structure of the magnetization.
To clarify how the local magnetization develops in the lattice
as the interaction strength increases beyond the critical interaction, 
we consider the IPR of the magnetization,
which quantifies the degree of localization of a nonuniform spatial pattern.
The corresponding quantity for the $n$th moment is given by
\begin{align}
  {\rm IPR}_n=\sum_i \left( \frac{|m_i|^n}{\sum_j |m_j|^n}\right)^2.
\end{align}
When the magnetization is spatially uniform throughout the system, the IPR takes the value $1/N$.
By contrast, if only one site is magnetized, the IPR becomes unity.
Therefore, this quantity is appropriate for discussing the spatial inhomogeneity
of the magnetization in the system.
Note that, in general, the IPR$_1$ and IPR$_2$ are often used to characterize localization.
Since the IPR$_n$ is related to the $n$th moment of the local magnetization,
the IPR$_2$ is more sensitive to the sharpness of the magnetization profile,
{\it i.e.}, the prominence of the magnetization peak.
By contrast, the IPR$_1$ captures the overall spatial extent of the magnetization distribution.

\begin{figure}[htb]
  \includegraphics[width=\linewidth]{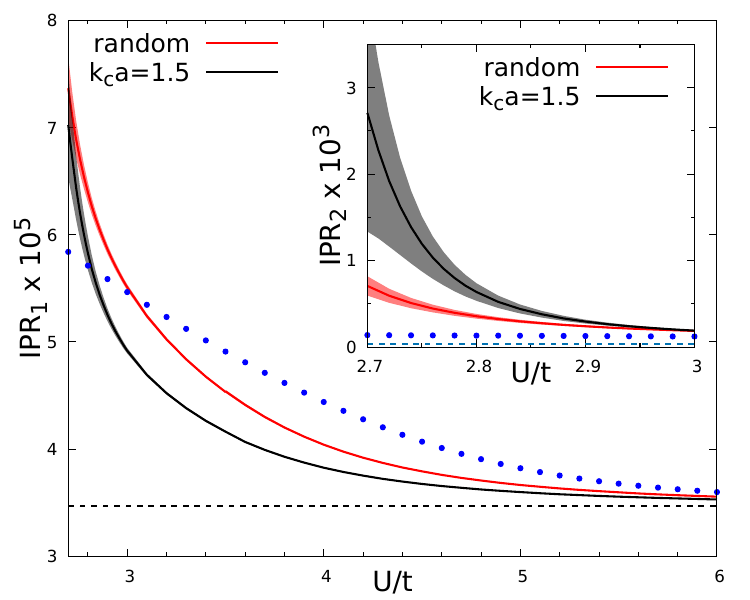}
  \caption{
    Red (black) line represents IPR$_1$ of the magnetizations
    as a function of $U/t$
    in the Hubbard models with random (stealthy hyperuniform) bond distribution.
    The system size is $N=2L_1L_2$ with $L_1=L_2=120$.
    Dotted line represents IPR$_1$ for the magnetization of the Hubbard model
    on the quasiperiodic bond distribution.
    Dashed line indicates the lower bound of the IPR of the system.
    Shaded areas are estimated by using the standard deviation of
    the results obtained from nine independent bond distributions.
    Inset shows IPR$_2$ for the same set of systems.
  }
  \label{ipr}
\end{figure}
Figure~\ref{ipr} shows the IPR of the magnetizations in the systems
with randomly disordered and stealthy hyperuniform bonds.
When $U/t\sim 6$,
the corresponding curves are almost identical for both IPR$_1$ and IPR$_2$,
which is consistent with the fact that the local
magnetizations have comparable magnitudes and similar
distributions (see Fig.~\ref{mag2}).
Furthermore, as the interaction strength increases, 
these values approach $1/N$ since 
the magnetization approaches a single uniform value (not shown).

\begin{figure}[htb]
  \includegraphics[width=\linewidth]{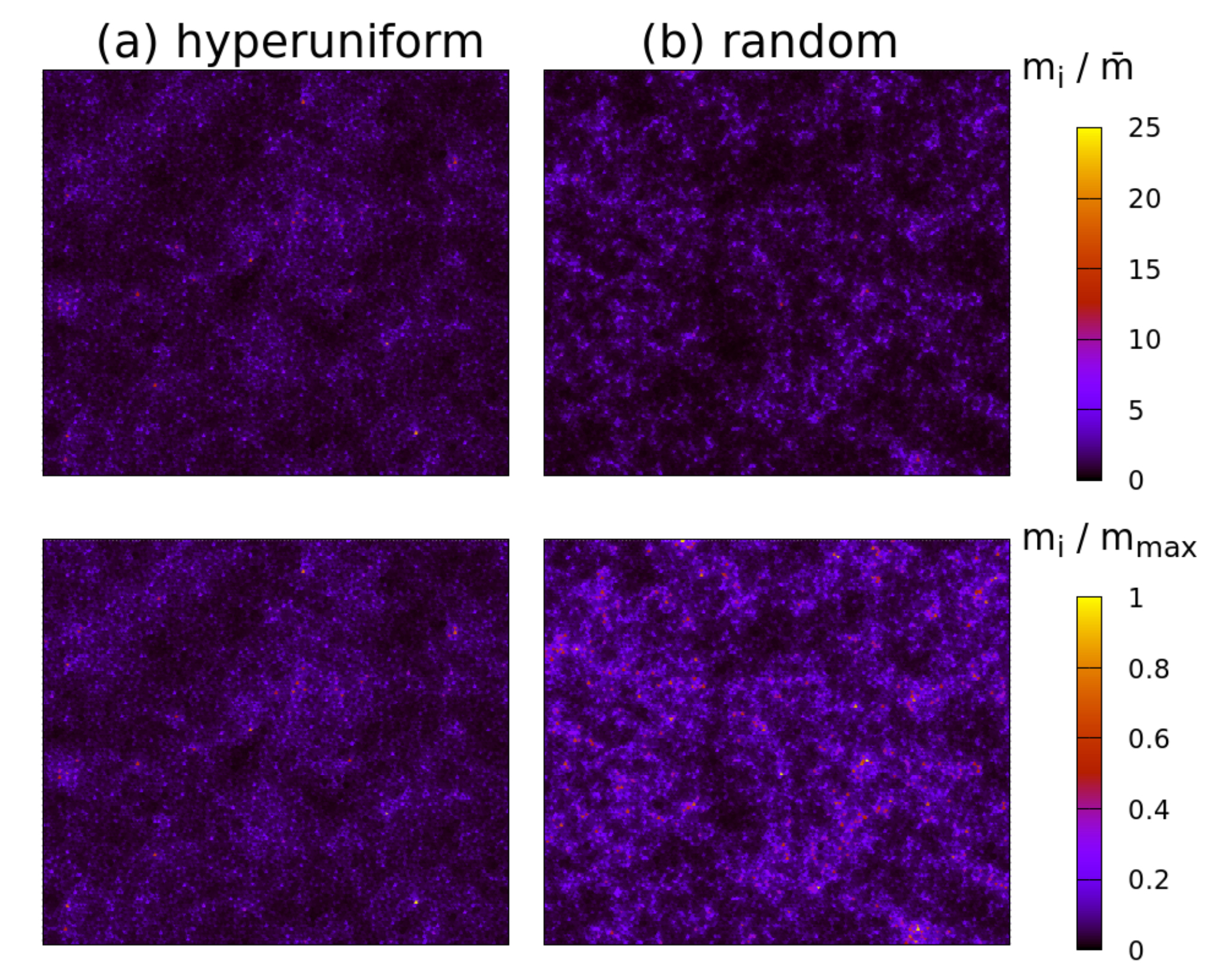}
  \caption{
    The spatial distributions of the moments
    in the Hubbard model with $U/t=2.8$ and $L_1=L_2=120$,
    where the hopping integrals are (a) stealthy hyperuniformly and
    (b) randomly disordered.
    Upper and lower panels show the results for $m_i/\bar{m}$ and $m_i/m_{max}$.
  }
  \label{dis}
\end{figure}

In contrast, at smaller interaction strengths $(U/t\sim 2.8)$, 
the IPR$_1$ of the stealthy hyperuniform system is slightly smaller than
that of the randomly disordered system, 
indicating a more spatially homogeneous magnetization in the stealthy hyperuniform case. 
This tendency may be visible in the upper panels of Fig.~\ref{dis}. 
By contrast, the IPR$_2$ exhibits the opposite behavior:
its value for the stealthy hyperuniform system is larger than that 
for the random case. 
This implies that, in the stealthy hyperuniform case, the magnetization is spatially more homogeneous, 
while only a limited number of sites host relatively large local magnetizations. 
The presence of such rare sites in the stealthy hyperuniform system has been confirmed 
by examining multiple independent realizations 
of the stealthy hyperuniform bond configurations
although the variance is large, as shown in Fig.~\ref{ipr}.
In the randomly disordered system, many sites instead exhibit magnetizations of comparable amplitude, 
leading to a smaller IPR$_2$.
This is clearly observed in the lower panels of Fig.~\ref{dis},
where the magnetization is normalized by its maximum. 
These results demonstrate that stealthy hyperuniformity fundamentally
changes the development of magnetic order, 
affecting not only the spatial distribution of the magnetization
but also the statistics of its local amplitudes.

We also compare the results for the above disordered systems with those for the quasiperiodic system,
where the same local structures are quasiperiodically distributed.
The IPR results are shown as the dotted lines in Fig.~\ref{ipr}.
For IPR$_1$, the value is slightly larger than those for the other disordered systems when $U/t\gtrsim 3$.
In contrast, IPR$_2$ takes a smaller value,
which is consistent with the fact that large magnetizations emerge
at the C$_3$ vertices,
where the fraction is larger than those for the disordered cases [see Fig.~\ref{fig:vertex}(b)],
and possess nearly identical amplitudes
due to the similar local structure.

The stability of the magnetically ordered state
depends not only on the presence of long-range correlations
in the bond distribution but also crucially on their specific nature.
In this study, we consider systems in which the number ratio of
$\alpha$ and $\beta$ bonds
is fixed to the golden mean, allowing for a direct comparison between quasiperiodic,
stealthy hyperuniform, and randomly disordered cases.
In the quasiperiodic system, a large number of C$_3$ sites are distributed
in a quasiperiodic and spatially correlated manner,
enabling them to act cooperatively.
As a result, the magnetically ordered phase
is stabilized at a smaller critical interaction $U_c$.
By contrast, in the stealthy hyperuniform system,
although bonds exhibit long-range correlations, C$_3$ sites are
extremely sparse and largely isolated,
including rare, strongly magnetized sites.
As a result, the value of IPR$_2$ becomes large,
reflecting the dominance of a small number of sites
with relatively large local magnetizations,
while the formation of global magnetic order is suppressed.
The randomly disordered system exhibits intermediate behavior.

In the regime $U \lesssim U_c$, most sites remain nonmagnetic,
while enhanced local magnetization appears only at a subset of sites.
Such behavior may be reminiscent of rare-region effects.
However, the present mean-field calculations do not allow us to establish
the existence of a Griffiths phase or Griffiths singularities~\cite{Griffiths}.
We therefore regard the possible connection to Griffiths-phase-like behavior
as a qualitative interpretation rather than a conclusive identification.

Since the stealthiness imposed on the bonds does not directly determine
that of the vertex distributions, 
we briefly comment on the spatial correlations of the vertex types
(C$_0$, C$_1$, C$_2$, and C$_3$).
The corresponding structure factors exhibit signatures of
stealthy hyperuniform correlations 
characterized by the cutoff wave vector $k_c$, 
but they do not satisfy the strict stealthy condition
that the structure factor vanishes for $k<k_c$ [see Fig.~\ref{vsq}(a)].
When stealthy hyperuniform constraints are imposed on both bonds and vertices, 
the fraction of C$_3$ vertices is strongly suppressed, as shown in Fig.~\ref{fig:vertex}(b), and 
the resulting point patterns, density of states, and magnetic properties
change only slightly.
The details are presented in Appendix~\ref{app}.
This suggests that, although global properties remain robust, 
the presence of such rare and spatially isolated sites may play a key role in the magnetic response.

\section{Summary}\label{sec:summary}
We have studied how spatial correlations in bond disorder affect the
electronic and magnetic properties of the Hubbard model on the honeycomb lattice.
Although the low-energy linear density of states remains robust,
hyperuniform correlations significantly modify
higher-energy electronic states, where the corresponding wave function
becomes extended.

Our real-space Hartree analysis has demonstrated that
the staggered magnetization in the antiferromagnetically ordered state is sensitive to
the stealth property of the bond distribution,
highlighting the importance of structural correlations beyond simple randomness.
Together with the comparisons to quasiperiodic lattices,
these results underscore hyperuniform disorder as an effective mechanism for
tuning electronic states and magnetic properties
in correlated systems.

Our calculations were restricted to the honeycomb lattice,
while the stealth properties do not depend on the details of the underlying lattice.
Since the honeycomb lattice is bipartite, we expect that the single-particle states and
the magnetic properties in the ordered state would behave
similarly on other bipartite lattices.
For non-bipartite lattices with triangular structures, however,
geometrical frustration may qualitatively change the single-particle states and magnetic properties,
and this remains to be clarified.
A systematic study of such lattices is left for future work.

\begin{acknowledgments}
  We would like to thank M. Imada for valuable discussions.
  Parts of the numerical calculations were performed
  in the supercomputing systems in ISSP, the University of Tokyo.
  This work was supported by Grant-in-Aid for Scientific
  Research from JSPS, KAKENHI Grants No. JP22K03525,
  JP25H01521, JP25H01398 (A.K.) and JP25H01397, JP25H01399 (T.S.).
\end{acknowledgments}

\appendix

\section{Constrained stealthy hyperuniform}\label{app}

\begin{figure*}[htb]
  \includegraphics[width=\linewidth]{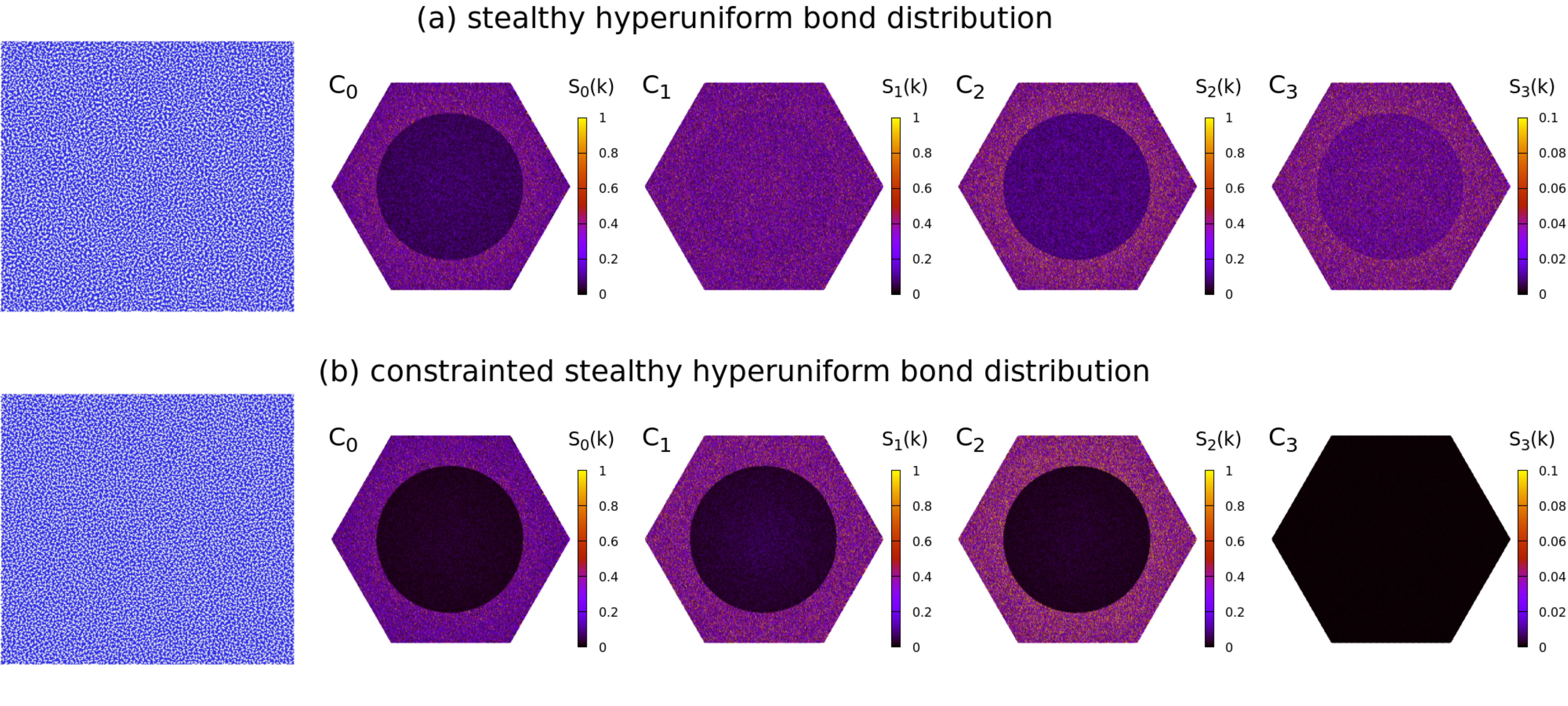}
  \caption{
    (a) Distribution of stealthy hyperuniform bonds with $k_ca=1.5$ and
    the corresponding structure factors for the four vertex types.
    (b) Same as (a), but with additional stealthy hyperuniform constraints
    imposed on both bonds and vertices.}
  \label{vsq}
\end{figure*}

In the main text, we have constructed stealthy hyperuniform bond configurations
and clarified that
the fractions of the four vertices change as $k_c$ increases
[see Fig.~\ref{fig:vertex}(b)].
In particular, 
the fraction of C$_3$ vertices which play an important role for magnetic properties,
is strongly suppressed.
To gain insight into the spatial correlations for the vertices, 
we analyze the structure factors $S_i({\bf k})$ for C$_i$ vertices.
Figure~\ref{vsq}(a) shows the structure factors corresponding to
the spatial distributions of the C$_0$, C$_1$, C$_2$, and C$_3$ vertices.
Signatures of stealthy hyperuniform correlation characterized
by the cutoff wave vector $k_c$ are visible in all cases
although they are less pronounced for the C$_1$ vertices.
However, the strict stealthy condition $S_i(k)=0$ for $k<k_c$ is not satisfied.
To examine the effect of enforcing this condition, we additionally impose stealthy hyperuniform constraints 
on the vertex distributions and construct bond configurations with an additional constraint.
To this end,
we introduce the cost function~\cite{Asakura2}
\begin{align}
  \Psi\left(\{{\bf r}^{(\alpha)}\}\right) =
  \Phi\left(\{{\bf r}^{(\alpha)}\}\right) + \sum_{{\bf k}i}V({\bf k})\Big[ S_i({\bf k}) -S_{\rm target}({\bf k})\Big]^2.
\end{align}
We optimize the bond configurations so that the cost function is minimized.
The bond configuration and its structure factor for $k_ca=1.5$ are shown in Fig.~\ref{vsq}(b).
The resulting structure factors satisfy the stealthy condition for each vertex type. 
Although the bond configurations are visually similar to those for
the simple stealthy hyperuniform system with $k_ca=1.5$,
the relative fractions of the different vertex types change significantly, as shown in Fig.~\ref{fig:vertex}(b).
Consequently, C$_3$ vertices are strongly suppressed,
consistent with the fact that $S_3({\bf k})\sim 0$ over the entire $k$-space.

\begin{figure*}[htb]
  \includegraphics[width=\linewidth]{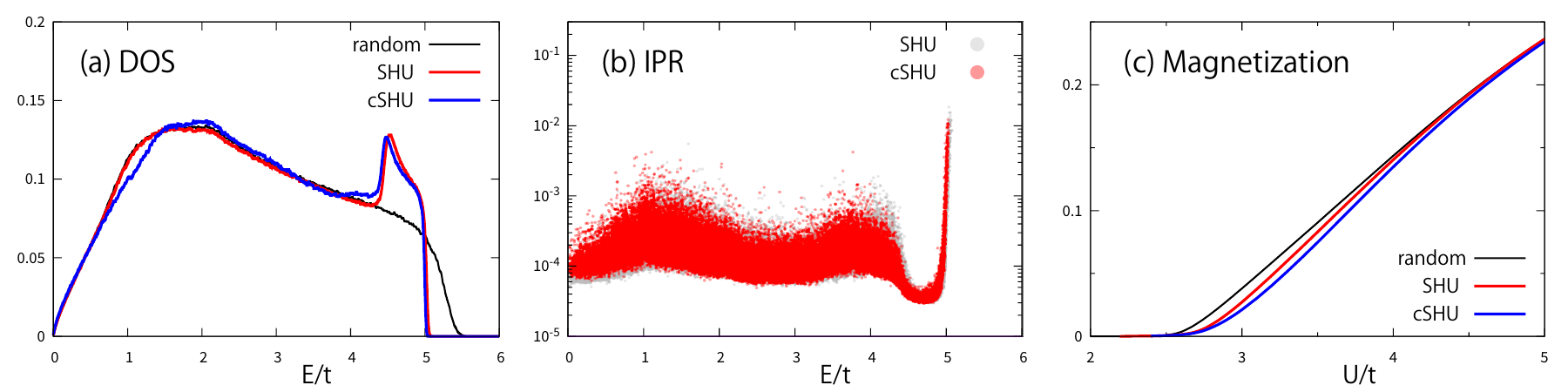}
  \caption{
    (a) DOS for the tight-binding model with $L_1=L_2=512$ and
    (b) IPR spectra for the model with $L_1=L_2=256$.
    (c) Average of local magnetizations as a function of the Coulomb interaction
    in the disordered systems with $L_1=L_2=120$.
  }
  \label{dosv}
\end{figure*}
To examine how the imposed stealthiness of the vertices affects
the electronic properties,
we diagonalize the tight-binding model on the disordered bonds.
The DOS and IPR spectra are shown in Figs.~\ref{dosv}(a) and (b).
Only slight differences are observed in these quantities.
The average magnetization in the Hubbard model is shown in Fig.~\ref{dosv}(c).
We find that the magnetization for the constrained stealthy hyperuniform case is always lower than
that for the SHU case.
This reduction is consistent with the suppressed fraction of the C$_3$ sites, as discussed in the main text.

\section{System size dependence of the IPR spectrum}\label{size}

\begin{figure}[htb]
  \includegraphics[width=\linewidth]{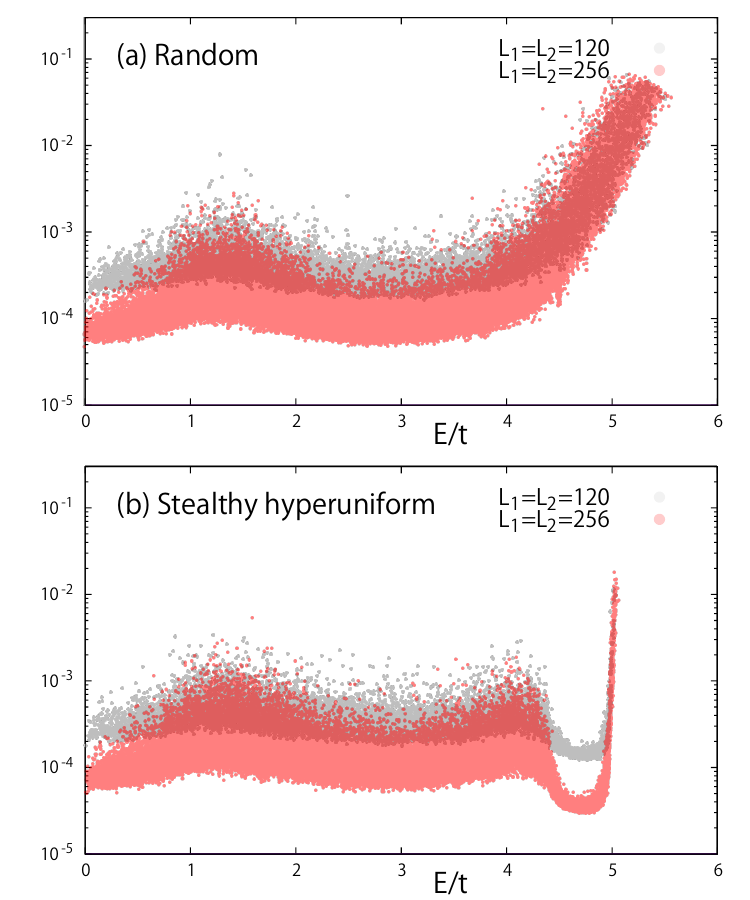}
  \caption{IPR spectra for the tight-binding model
    with (a) random and (b) stealthy hyperuniform bond distributions
    for the system size $L_1=L_2=120$ (gray) and $256$ (red).
  }
  \label{ipr4a}
\end{figure}

We here demonstrate the system size dependence of the IPR spectrum.
It is known that the IPR spectrum is independent of the system size for the localized states,
while it is proportional to $1/N$ for the extended states.
This allows us to distinguish between extended and localized eigenstates.
Figure~\ref{ipr4a} shows the IPR spectra for the tight-binding model with
random and stealthy hyperuniform bond distributions.
Away from the band edges, the IPR clearly decreases with increasing system size in both cases,
indicating that the extended states are realized.
On the other hand, the IPR hardly depends on the system size for $|E|/t>4.5$ for the random case and
for $|E|/t\sim 5$ for the stealthy hyperuniform case,
indicating that the localized states are realized around the band edges.
In particular, the eigenstates in the range $4.5 \lesssim |E|/t \lesssim 4.9$,
which are localized in the random case,
exhibit a clear system size dependence in the stealthy hyperuniform case.
These results indicate that the stealthy hyperuniform bond disorder delocalizes
the eigenstates near the band edge.

\bibliographystyle{jpsj}
\bibliography{./refs}

\end{document}